# An Improved Laboratory-Based XAFS and XES Spectrometer for Analytical Applications in Materials Chemistry Research

Evan P. Jahrman[1], William M. Holden[1], Alex S. Ditter[1,2], Devon R. Mortensen[1,3], Gerald T. Seidler[1] (*), Timothy T. Fister[4], Stosh A. Kozimor[2], Louis F.J. Piper,[5] Jatinkumar Rana,[5] Neil C. Hyatt,[6] and Martin C. Stennett[6]

[1]Physics Department, University of Washington, Seattle, WA  98195-1560
[2]Chemistry Division, Los Alamos National Laboratory, Los Alamos, NM 87545
[3]easyXAFS LLC, Seattle, WA 98122
[4]Chemical Sciences and Engineering Division, Argonne National Laboratory, Lemont, IL 60439
[5] Department of Physics, Binghamton University, Binghamton, NY 13902
[6] Materials Science and Engineering Dept., The University of Sheffield, Mapping Street, Sheffield, S1 3JD, UK

**ABSTRACT**

X-ray absorption fine structure (XAFS) and x-ray emission spectroscopy (XES) are advanced x-ray spectroscopies that impact a wide range of disciplines.  However, unlike the majority of other spectroscopic methods, XAFS and XES are accompanied by an unusual access model, wherein; the dominant use of the technique is for premier research studies at world-class facilities, i.e., synchrotron x-ray light sources.  In this paper we report the design and performance of an improved spectrometer XAFS and XES based on the general conceptual design of Seidler, *et al*., Rev. Sci. Instrum. 2014.  New developments include reduced mechanical degrees of freedom, much-increased flux, and a wider Bragg angle range to enable extended x-ray absorption fine structure (EXAFS) for the first time with this type of modern laboratory XAFS configuration. This instrument enables a new class of routine applications that are incompatible with the mission and access model of the synchrotron light sources.  To illustrate this, we provide numerous examples of x-ray absorption near edge structure (XANES), EXAFS, and XES results for a variety of problems and energy ranges.  Highlights include XAFS and XES measurements of battery electrode materials, EXAFS of Ni and V with full modeling of results to validate monochromator performance, valence-to-core XES for 3d transition metal compounds, and uranium XANES and XES for different oxidation states.   Taken *en masse*, these results further support the growing perspective that modern laboratory-based XAFS and XES have the potential to develop a new branch of analytical chemistry.

## I. Introduction

X-ray absorption fine structure (XAFS) analysis is an especially capable and impactful tool for interrogating a material's local electronic and atomic structure. This element-specific technique encompasses both the x-ray absorption near edge structure (XANES), an acutely sensitive probe of a compound's oxidation state and molecular geometry, and the extended x-ray absorption fine structure (EXAFS), which is routinely used to extract multi-shell coordination numbers and bond lengths. These techniques enable premier scientific research campaigns in catalysis,[1-2] energy storage,[3-4] actinide chemistry,[5-7] heavy metal speciation in the environment,[8-10] etc. Likewise, the partner process, x-ray emission spectroscopy (XES), has been used to assess spin and ligand character, notably in critical discoveries of magnetic phase transitions under geophysical conditions.[11-12] At present, XES continues to emerge as an important measure of valence-level (occupied) electronic state properties through improved theoretical treatment of the valence-to-core (VTC) and core-to-core (CTC) XES. However, as has been pointed out several times in the modern history of XAFS and XES, and most recently by Seidler,[13] these x-ray spectroscopies suffer from an anomalous access model. In general, XAFS and XES studies require access to synchrotron facilities with entry requirements that limit more introductory, routine, or high-throughput analytical studies that, by contrast, are common for NMR, XPS, or optical spectroscopies where high-access benchtop equipment is easily available.

Over the last several decades, the capabilities of lab-based XAFS and XES instruments have rapidly grown. Researchers now report spectrometers operating as low as the C K-edge (284 eV)[14] using a laser-produced plasma source. Other spectrometers probe the S and P K emission lines (~ 2-2.5 keV) using double crystal monochromators,[15-17] a dispersive Rowland circle geometry,[18-21] and an instrument in the von Hamos geometry.[22] A variety of von Hamos instruments exist which are intended to operate in the ~3-12 keV range needed for studies of first row transition metals and lanthanides.[23-26] Many spectrometers operating in this range can be directly integrated into synchrotron beamlines.[27] For similar energies, a large number of XAFS spectrometers employing a Rowland circle geometry exist.[28-33] Lastly, higher energies, including the Au Kβ (78 keV), are accessible via Laue-type spectrometers.[34] We focus here on the case of Rowland circle geometries with a spherically bent crystal analyzer (SBCA), which has been extensively developed by some of the present authors.[13, 35-40]

The purpose of the present manuscript is to describe the design and performance of what is our latest-generation of improvements upon the first prototype instrument using an SBCA.[13] These are embodied in two nearly identical spectrometers, one at the University of Washington (UW) in Seattle and one at Los Alamos National Laboratory (LANL). Each of these sites is more than 1000 km away from the nearest synchrotron x-ray light source. The spectrometer improvements include several simplifications to the monochromator mechanical system that decrease its operation from five to only two motorized degrees of freedom and the selection of a ten-fold higher power x-ray tube that retains the small size and necessary anode characteristics to meet the needs of laboratory XAFS and XES.

The manuscript continues as follows: First, in section II, we describe the new monochromator. Important highlights include decreased mechanical complexity of the new design and modification of the drive configuration to increase its Bragg angle range while minimizing its air-absorption path and overall footprint. Second, in Section III we present and discuss results for XANES and EXAFS of several materials reflecting contemporary interest in materials chemistry and other specialties. Examples include reference metal foils, battery electrode laminates of several different compositions, a family of reference Ce compounds, and uranium-rich materials. In all cases we find good agreement with prior synchrotron studies. For the recorded EXAFS spectra on the metal foils, we further present a full Fourier-transform analysis using standard methods, and again find high quality results. Next, in Section VI we present and discuss results for XES from a wide variety of elements, chemical systems, and emission lines. This includes both deep-shell emission lines and the VTC XES that provides direct insight into chemical bonding. In sections III and IV, care is taken to provide measurement times, thus serving as useful benchmarks for assessing the feasibility of future studies using SBCA-based laboratory monochromators.

## II. Experimental
## II.A. Monochromator Design

Throughout the period between first publication[13] and this work, several advances in the spectrometer design have been made. Specifically, we have integrated a higher powered x-ray source, rotated and greatly elongated the source and detector stages, implemented passive tracking of the SBCA position (removing a motorized degree of freedom), and enacted the tiltless optic alignment introduced by Mortensen *et al*.[39] (removing two additional motorized degrees of freedom). These changes were motivated by a focus on greater count rates, instrument stability, ease-of-use, and achieving a wider useful energy range with each analyzer crystal orientation.

An overview of the new spectrometer design is given in Fig. 1. The approach uses linear translation stages to generate fine rotations (Bragg angle steps) and 'steering bars' to maintain alignment between the source, detector, and SBCA. This design was based directly on our prototype system.[13] We now summarize similarities and differences of the two new instruments with respect to the prototype instrument.

First, the prototype system used a motorized translation stage underneath the SBCA to maintain its position on the 1-m Rowland circle while the source and detector positions (and hence Bragg angles) were scanned. In the present instrumentation, a passive linear translation stage with two carriages is used; one carriage for the SBCA and another carriage for a pin located at the moving center of the Rowland circle. Coupling bars with lengths equal to the radius of the Rowland circle constrain the Rowland-center pin to be the correct distance (0.5 m) from pins underneath both the SBCA and the source location. This direct mechanical coupling provides exceptional scan-to-scan reproducibility and decreases instrument complexity by removing one motorized degree of freedom.

Second, the source and detector stages have been rotated and made much longer than in the earlier system. The longer travel range allows access to Bragg angles between 55 and 85 degrees, a considerable improvement over the prototype that allows a much wider energy range for each crystal orientation of SBCA. This change decreases the total number of SBCA optics required of XAFS and XES analysis. Moreover, it extends the utility of the spectrometer beyond XANES, enabling EXAFS studies for several elements. The stage rotation requires some careful comment. The resulting stage geometry is shown in Fig. 1 and the rotation parameter $\alpha$ is defined in Fig. 2a. The issue that motivates the rotation of the stages is the desire to minimize the linear travel of the SBCA needed to maintain its position on the (traveling) 1-m Rowland circle. Long SBCA travel is not mechanically onerous, but clearance is required with respect to the helium box (not shown) enclosing the beampath to reduce air absorption. (not shown). When the SBCA has a long travel, the helium box must be made shorter which results in higher air absorption for most operations. To address this problem, a suitable value of $\alpha$ can be deduced from geometric considerations. As the source and detector are swept outward to smaller Bragg angles, the SBCA is necessarily displaced to ensure the source and detector remain on the Rowland circle. This displacement $d(\theta_B)$ is given by

$$d(\theta_B) = -R * sec[\alpha] * (cos[\theta_o - \alpha] + cos[\alpha + 2\theta_B]), \qquad (1)$$

where $R$ is the radius of the Rowland circle, $\theta_B$ is the Bragg angle, and the displacement is measured relative to the position of the SBCA when $\theta_B = 85°$, this value is denoted above as $\theta_o$. In Figure 2b, Eq. 1 is plotted as a function of $\theta_B$ for various values of α. It is apparent that translation of the SBCA, and consequently attenuation due to air outside of a fixed helium enclosure, can be minimized by an appropriate choice of α. This translation is minimized when the SBCA's travel is symmetric across the angle range, which can be enforced by choosing α to be equal to 180° minus twice the midpoint of the angle range. For a $\theta_B$ range of 85° to 55° the SBCA's displacement is minimized when α = 40°, as is utilized in Figure 1. Moreover, an additional benefit of the stage rotations is a smaller instrument footprint.

Third, the present design discontinues the traditional two-axis tilt alignment of the SBCA in favor of orienting the crystal miscut into the plane of the source and detector and enforcing a constant angular offset of the detector, as described by Mortensen and Seidler.[39] This removes two motorized degrees of freedom and also enables easy, reproducible exchange of different SBCAs for different energy ranges. Here, SBCAs are aligned by performing repeated detector scans at different rotations of the optic about its natural cylindrical axis. This fast process gives a permanent alignment orientation. See Fig. 3 for representative calibration scans. Note that the highest count rates are generally observed when the crystal miscut is oriented into the Rowland plane, as the SBCA is rotated in either direction away from this orientation, the centroid of the corresponding scans move in the same direction away from the peak at optimal orientation.

Fourth, from a practical standpoint, the primary benefit of a high-flux source is shorter acquisition times and thus higher potential instrument throughput. Moreover, greater flux can broaden an instrument's limit-of-detection, thus enabling studies of particularly dilute samples or weak transition lines. Nonetheless, there exist several points of concern when utilizing a high-powered tube source. Namely, high-powered sources typically increase demands on cooling, require progressively more expensive components and high voltage supplies, and their size typically scales in a nontrivial way with increasing tube power, thus posing a challenge toward integration into the synchronous scanning motif. To balance these considerations, the Varex

VF50 and VF80 x-ray tube sources are used in the present instruments. These tube sources operate at 50 and 100 W, respectively, and we use either W or Pd anodes as needed to optimize signal levels or avoid tube-source fluorescence lines. These x-ray tubes have small spot sizes (0.5 – 1 mm) and use a 90-degree take-off geometry giving especially efficient generation of x-ray flux per unit electron beam power. By comparison, the x-ray tubes used in the prototype spectrometer[13] were 5-12 W total power and used transmission-geometry anodes with ~3x lower efficiency per unit electron beam power due to absorption during transit through the anode material itself.

Fifth, the high voltage supplies (Spellman uX at LANL and both uX and uXHP at UW) are factory customized to not exceed 35 kV accelerating potential. This has little effect on the final flux generated at useful energies but has the considerable advantage that the radiation enclosure can then be made from 3.175 mm steel. The new radiation enclosure is a welded steel box with two gas-spring loaded top-facing doors for access to the spectrometer. Two labyrinths provide pass-throughs for cables and gas flow lines.

Finally, at UW the same silicon drift detector (Amptek X-123 SDD) is used as in the prototype spectrometer, while a PIN diode is used in the spectrometer at LANL (Amptek X-123 Si-PIN). Here, however, the ~4-mm active height of each detector does prove somewhat limiting. As $\theta_B$ deviates strongly from backscatter (decreasing $\theta_B$) the height of the refocused beam on the detector quickly becomes taller than the active height of the detector, resulting in significant inefficiency. As discussed below, and in more detail in an upcoming manuscript,[41] this can be ameliorated by incorporation of a taller SDD or by use of a toroidal curved crystal analyzer that is tailored to the Bragg angle range of interest.

**II.B. Sample Preparation Details**

In the results and discussion below, we present numerous studies of both XAFS and XES for a wide range of materials. In this subsection we briefly summarize the preparation or provenance of each system.

$CeO_2$ was prepared by thermal decomposition of cerium (IV) oxalate, $Ce(C_2O_4)_2 \cdot xH_2O$ at 800 °C, for 1h, in air, as described by Stennett *et al*.[42] $CePO_4$ (with the monazite structure) was prepared by solid state reaction of stoichiometric quantities of $CeO_2$ and $NH_4H_2PO_4$: an intimate mixture of these reagents, prepared by hand grinding with a mortar and pestle, was heated at 1100 °C, for 8h, in air. Analysis of the products by powder X-ray diffraction confirmed the synthesis of single phase materials. Specimens were prepared for XAS analysis by sieving to less than 63 μm before mixing with polyethylene glycol and pressing into 13-mm diameter pellets having suitable $\mu x$ for transmission-mode study.

ε-VOPO$_4$ investigated in the present study was prepared by hydrothermal synthesis.[43] Thin laminates for XAS investigation were prepared by mixing ε-VOPO$_4$ powder with graphene and polyvinylidene fluorine (PVDF) as binder in a weight ratio 75:15:10 using 1-methyl-2-pyrrolidinine (NMP) as the solvent. The resultant slurry was tape cast onto an aluminum foil and dried in air at 60 °C. Circular discs of about 13mm diameter were punched out of the coated aluminum foil and sealed between the adhesive-coated Kapton tapes.

Commercial nickel- manganese- cobalt- (NMC) oxide battery cathode laminates were manufactured in a 6:2:2 stoichiometric ratio between the transition metals. The cathode laminate was cast with a 5 wt % PVDF binder and carbon on a 10 μm thick aluminum current collector. Laminates were prepared in two states of charge, a pristine uncharged laminate and a charged laminate harvested from a coin cell. The latter was sealed in an aluminum-coated polyimide envelope during measurement to reduce interaction with air.

Uranium(IV) hexachloride, (PPh$_4$)$_2$UCl$_6$, was prepared as previously described.[44] Uranyl tetrachloride was prepared in a modified version of previous syntheses.[45-46] This involved addition of two equivalents of tetramethyl ammonium chloride (NMe$_4$Cl) to UO$_2^{2+}$ in concentrated hydrochloric acid (HCl, 12 M). Within 1 week crystals formed and the compounds identity was then confirmed by single crystal X-ray diffraction. ***Caution!** The $^{238}$U isotope is a low specific-activity α-emitting radionucleotide and its use presents a hazard to human health. This research was conducted in a radiological facility with appropriate analyses of these hazards and implementation of controls for the safe handling and manipulation of these toxic and radioactive materials.* Samples were prepared by grinding the (PPh$_4$)$_2$UCl$_6$ (20 mg) with boron nitride (BN, 40 mg) for two minutes. An aluminum spacer with interior dimensions 1 mm x 5 mm x 20 mm was filled with the resulting powder and sealed in two layers of polyimide tape.

Finally, metal foils were acquired from EXAFS Materials. These include a 6 μm Ni foil, a 5 μm V foil, a -400 mesh Mn foil, and a 25 μm Y foil. Also, the V$_2$O$_3$, VO$_2$, V$_2$O$_5$, NaAsO$_2$, and Na$_2$HAsO$_4$·7H$_2$O powders measured in XES were purchased from commercial vendors.

**II.C. Synchrotron XAS Measurement Details**

XAS measurements of ε-VOPO4 were carried out at the beamline 9-BM of Advanced Photon Source (APS) in USA. Data were collected in the transmission mode at the V K-edge using the Si (111) double-crystal monochromator, which was slightly detuned to suppress higher harmonics. Absolute energy calibration of the monochromator was carried out by measuring the reference foil of pure V simultaneously with the sample. Intensities of the incident beam and the beams transmitted through the sample and the reference foil were recorded using the gas-filled ionization chambers. All spectra were energy-calibrated with respect to the first peak in the derivative spectrum of pure V. Data processing operations were carried out using the software ATHENA of the package IFEFFIT.[47]

Ce $L_3$ edge XAS data of $CeO_2$ and $CePO_4$ (with the monazite structure) were acquired on beamline X23-A2 of the National Synchrotron Light Source (NSLS), Brookhaven National Laboratory (BNL), USA. This beamline is configured with a piezo-feedback stabilized Si (311) upwards reflecting monochromator and single bounce harmonic rejection mirror. Data were acquired in transmission mode using finely ground specimens dispersed in polyethylene glycol (PEG) to achieve a thickness of one absorption length. 0.5 eV steps were used over the absorption edge with a dwell time of 5 seconds per point. Incident and transmitted beam intensities were measured using ionization chambers operated in a stable region of their I/V curve filled with mixtures of He and Ar or $N_2$.

The U $L_3$ x-ray absorption near edge spectra (XANES) were measured at the Stanford Synchrotron Radiation Lightsource (SSRL), under dedicated operating conditions (3.0 GeV, 5%, 500 mA using continuous topoff injections) on end station 11-2. This beamline was equipped with a 26-pole, 2.0 tesla wiggler, utilized a liquid nitrogen-cooled double-crystal Si (220) monochromator, and employed collimating and focusing mirrors. A single energy was selected from the white beam with a liquid-$N_2$-cooled double-crystal monochromator utilizing Si (220) ($\varphi$ = 0) crystals. Although the crystals were run fully-tuned, higher harmonics from the monochromatic light were removed using a 370 mm Rh coated harmonic rejection mirror. The Rh coating was 50 nm with 20 nm seed coating and the substrate was Zerodur. Vertical acceptance was controlled by slits positioned before the monochromator. The harmonic rejection cut-off was set by the mirror angle, thereby controlling which photons experience total external reflection. The horizontal and vertical slit sizes were 15 mm and 1 mm.

**III. XAFS and EXAFS Results and Discussion**

### III.A. Basic Instrument Performance

The present instrumentation was evaluated according to several performance criteria, including typical count rates. In Figure 4, the intensity of the incident beam in an absorption configuration was measured across the full angular range of the instrument. Near backscatter ($\theta_B$ = 90 °), count rates near 50,000/s are observed at 100 W x-ray tube power with the Pd anode (this and all subsequent measurements were performed at 35 kV accelerating potential), yet the count rate quickly drops to around 15 % of this value at especially low Bragg angles. The reason for this decline is the limited size of the detector. In the present design, x-rays are refocused to a line at the detector due to sagittal error. The height of this line increases as the source and detector travel to lower Bragg angles and only a portion of this line is measured as permitted by the finite size of the detector. The decline in count rates observed in Figure 4 is consistent with ray tracing simulations reported elsewhere.[41] If a larger detector or toroidal optic is integrated into the design, consistent count rates could be observed across the instrument's angular range.[41]

The spectrometer's reliability was assessed according to its scan-to-scan reproducibility. Figure 5 shows a series of consecutive scans collected in a transmission mode XANES configuration across the Mn K-edge of the Mn foil. Also shown is the residual of each scan with respect to the first and an envelope of two standard deviations as calculated from the incident flux by Poisson statistics. The residuals are well captured by Poisson statistics.

The monochromator performs well in a final, critical performance metric, its energy resolution. This parameter was assessed by measuring the XANES spectrum of a V metal foil. From Figure 6a, the laboratory-based instrumentation produces spectra nearly identical to those acquired at the synchrotron, however, minor changes in resolution can be observed by magnifying the especially sharp pre-edge feature found in this system. Convolving the synchrotron spectrum with a 0.4 eV FWHM Gaussian yields excellent agreement with the first set of laboratory-based measurements. One contribution to the broadening is that although the V K-edge is located at a Bragg angle of 79.2 ° for the Ge (422) optic, the spectrum will still exhibit some broadening due to Johann error. To investigate this effect, the outer portion of the SBCA was blocked with a Pb mask to produce a spectrum that is now broadened by only 0.2 eV relative to the spectrum reported by NSLS X23A2.

### III.B. XAFS Demonstration Studies

Here, we present the results of several XAFS studies using the lab spectrometer. These include XANES of battery materials, lanthanide and actinide compounds and also EXAFS of reference metal foils. The times for scan acquisitions of all demonstration studies are summarized in Table 1. Taken *en masse*, the results strongly support the usefulness of the lab spectrometer for a very wide range of concentrated systems where transmission-mode studies are possible. We begin with XANES.

First, electrical energy storage is a particularly promising application for laboratory-based x-ray spectroscopies.[48-51] Here, XANES is already established as a useful tool for the study of electronic properties at various levels of detail. For example, a routine approach utilizes XANES to assess the redox reversibility of battery materials during cycling.[52] Similarly, many examples exist of x-ray spectroscopies addressing more complex speciation inquiries, including lithiation dynamics in nickel cobalt aluminum oxide cathode materials,[53] discernment of the soluble Mn ion in a Li-Mn spinel electrode,[4] and evaluation of sulfide precipitation and under-utilization of active material as competing hypotheses for sub-optimal capacities in lithium sulfur batteries.[3]

Several other factors suggest lithium ion battery (LIB) cathode materials as an ideal system for laboratory-based x-ray instrumentation. Most importantly, the typical thickness of the metal oxide layer found on a cathode frequently gives edge steps $\Delta\mu \cdot x \sim 1 - 2$, as is desirable for XAS studies.[54] Also, the electrochemically active elements in modern LIB cathodes are often $3d$ transition metals, for which the K-edges are at energies high enough so that some air attenuation can be tolerated but low enough that the SBCA and other Bragg-based analyzers still have good efficiency.

The XANES spectra of two archetypal Li-ion battery materials, ε-VOPO$_4$ and NMC oxide laminates, are presented in Fig. 7 and Fig. 8a. The agreement between lab-based and synchrotron spectra in Fig. 7 is excellent, including the details of the pre-edge feature which is important for elucidating the molecular symmetry at the metal center.[55] Figure 8a presents NMC electrodes at two different states of charge. The charged and uncharged laminates exhibit multiple differences, including a pronounced shift in the edge position of the two systems. Such an edge shift is traditionally attributed to a change in oxidation state[56] and, in the present case, confirms the instrument's capability for element-specific tracking of redox behavior in cathode materials.

Moving away from the 3d transition metals, it is useful to next discuss lanthanide compounds. The L3-edges of the lanthanides are in a very similar energy range as the 3d transition metals, strongly suggesting good performance for our system, and there exists a large body of research using the dependence of XAFS spectral features on the speciation of lanthanide compounds.[57-59] Sample applications include high temperature, *in situ* analysis of: ceria-based oxide materials used in the activation and storage of oxygen,[60] the effect of annealing temperature on the valence state of cerium oxide nanoparticles manufactured to catalyze the oxidation of organic compounds or reduction of heavy metals in industrial waste streams,[61] and the mechanism by which cerium-containing films inhibit the corrosion of aluminum.[62]

XANES spectra of $CePO_4$ and $CeO_2$ taken in the lab are presented in Fig. 9a. The lab-based spectra are energy corrected by alignment with the reference cerium dioxide spectrum found in Hephaestus.[47] In particular, note that Poisson errors observed in Fig. 9a are far from eclipsing the shape of spectral features and that scan acquisition times, just as with the above transition metal study, are reasonable for many applications involving routine analytical characterization. This is true throughout the energy range from 6 keV to as high as the actinide $L_3$ edges at and above 17 keV, as we now show. The difference in height of the $CeO_2$ near-edge peak may be due to different preparation of the samples, as oxygen deficiency can commonly influence that feature.

Next, we address the high-energy range for applications of the laboratory spectrometer. While the present instrument design is not optimal for operation at 17 keV and beyond, it has proven quite effective. U $L_3$-edge XANES spectra for $(PPh_4)_2UCl_6$ is presented in Fig. 9b and directly compared to a synchrotron-based measurement. These measurements used the older, 50 W x-ray tube in the spectrometer at LANL. Clearly, U $L_3$ XANES can be measured in useful study times with our spectrometer; comparable results with a spectrometer of similar design have recently been reported by Bès, *et al*.[63]

The decreased performance of the lab spectrometers at high photon energy is due to limitations in the source, Bragg optic, and detector. The bremsstrahlung spectrum from the tube has the usual $\sim 1/E$ roll-off at high energy. This is complicated here, however, by our choice to hardware-limit the high-voltage supply to 35 kV, resulting in somewhat less proportional generation of ~17 keV photons than would be the case with a higher accelerating potential and the same total electron beam power. Combined with the narrower Darwin width of the SBCA for higher order reflections, the integral reflectivity is greatly decreased at higher photon energy.[64] There are some studies of higher-energy XAFS using laboratory-based instruments having Laue-style analyzers where the optic has much higher integral reflectivity from lower-order reflections, but where the effective solid angle is typically much reduced.[34, 65] The present detector also limits the efficiency for two reasons: it has only ~50 % quantum efficiency at these energies, and the active region of the detector used in the actinide study was only ~5-mm tall, so that ~2x flux was lost because of the vertical extent of the beam. Hence, the corresponding obvious upgrades to a 100-W x-ray tube and a taller detector with higher quantum efficiency will yield ~8x improved count rates on the same monochromator. The question of optimum lab-spectrometer design for high-energy XANES is very much an open question that could have high impact in heavy element chemistry (via L-edges) and 4$d$-chemistry (via K-edges).

The above studies demonstrate the broad versatility of the lab-based system for XANES studies. The extended oscillations pose a more stringent challenge, due to both the limitations imposed by Poisson statistics and the requirement of correct monochromator function over a wider energy range.

Two model systems were chosen to assess the present instrument's EXAFS capabilities relative to a synchrotron. These systems were nickel and vanadium metal, which are face centered and body centered cubic, respectively. The forward Fourier transform of metallic V or Ni's EXAFS spectrum was performed for photoelectron momentum $k = 3$ Å$^{-1}$ to $k = 10$ Å$^{-1}$ or $k = 12$ Å$^{-1}$, respectively. An isotropic expansion model was used for both systems and distinct Debye-Waller (DW) factors were assigned to the single scattering path associated with each neighboring atom. DW factors for collinear paths were calculated in the manner of Hudson *et al*.[66] while triangular paths were approximated from the single scattering path's DW factors. Fits were performed in Artemis[47] from R=1 to R=5.5 Å and included all scattering paths in that range. Resulting R-factors reveal the spectra to be well described by the fitted model. Similarly, the passive reduction factor, which is subject to inconsistencies according to individual beamline characteristics,[67] is within the range of values typically reported for robust fits. In Figures 10 and 11, excellent agreement is found between the lab and synchrotron-based measurements, as well as between experimental results and fitted models. Likewise, the physical quantities produced by the EXAFS fits are presented in Table 2 and Table 3, revealing excellent agreement between data acquired at different instruments and, in the case of Ni, with previously reported values.

**III.C. XES Demonstration Studies**

X-ray emission spectroscopy (XES) is seeing rapid growth as both a complement to XANES and as an emergent technique in its own right. Its sensitivity to the occupied local electronic density of states can often aid in assessing the oxidation state, spin state, covalency, state of protonation, or ligand environment of a given metal atom.[68-70]

From an experimental perspective, XES benefits from several pragmatic advantages in the laboratory environment. While the simpler sample preparation for XES than for transmission-mode XAFS is often relevant, the dominant issue is the efficient use of the incident x-ray flux. Conventional x-ray tubes are inherently broadband, showing a few strong fluorescence lines on top of a bremsstrahlung background. Monochromatizing the raw tube spectrum with a crystal analyzer selects only a modest solid angle of the total tube emission and also a tiny slice of the entire tube energy spectrum, decreasing broadband, wide-angle fluxes of $\sim 10^{13}$ /s or more to only $10^4 - 10^5$/s. However, direct illumination of the sample, as in non-resonant XES, utilizes a large solid angle and makes every incident photon above the relevant binding energy capable of stimulating the creation of a core-hole. Accordingly, a recent publication by the authors demonstrated lab-based XES measurements as a viable route to quantitatively assess metal speciation even in very dilute systems,[40] even with the very low powered x-ray tube of the earlier prototype spectrometer.[13]

Here, we present the results of several XES studies using the lab spectrometer. Thematically, the XES results are presented from lowest to highest energy emission lines. This begins with an overview of the Kβ lines for a collection of vanadium compounds including metallic vanadium, a suite of vanadium oxides, and vanadyl phosphate, a candidate material for energy storage applications. Similarly, routine valence-to-core (VTC) XES measurements of assorted zinc compounds sampling a variety of ligand environments are discussed. Next, arsenic Kα XES results suggest the potential of the present instrumentation for speciation studies of dilute environmental samples. Finally, less standard measurements of actinide L emission lines are presented. Note that, again, the acquisition times for all studies are summarized in Table 1 and repeated in the figure captions.

First, a range of chemical information is accessible in the V Kβ spectra in Fig. 12a. For example, note that the wide-energy range accessible by a single scan permits careful branching ratio studies of vanadium oxide $K\beta_{1,3}$ and $K\beta_{2,5}$ features which, as can be seen in Fig. 12a, are exceedingly well resolved.    An additional advantage of this range is the feasibility of robustly subtracting the tail of the main $K\beta$ emission from the VTC region to aid the analysis of the latter. Furthermore, Fig. 12b shows the $K\beta_{1,3}$ of a variety of vanadium oxide moieties, with a clear evolution in the spectrum as oxidation state changes.  Likewise, the $K\beta'$, which is split from the $K\beta_{1,3}$ by (3p,3d) exchange, can be seen to vary in intensity across the oxides.  For transition metals, the intensity of this feature often correlates with the number of unpaired 3*d* electrons and thus provides a measure of the spin state of the probed atom.  For some systems, the dependence of the $K\beta_{1,3}$ emission's energy on oxidation state can be muted, as is the case for Ni oxides.[71-72] However, it can be seen in Fig. 8b that there is a small but measurable shift between the $K\beta_{1,3}$ XES of two NMC laminates at different states of charge.  These spectra are also intense, allowing acquisition times on the order of minutes.

The VTC region for some Zn compounds is presented in Fig. 13.  The significance of this region warrants some discussion. In recent years, VTC-XES has emerged as a highly useful tool for the interrogation of a system's local electronic structure.  This method permits a direct probe of the orbitals involved in chemical bonding, and, as a result, is highly-sensitive to changes in oxidation state, covalency, state of protonation, and coordination environment.  As a unique case in point, VTC-XES is sometimes able to discern which of several light elements is ligated to a central metal atom, with considerable impact.  In 2002, Einsle *et al.* reported the presence of carbon, oxygen, or nitrogen as a central atom in iron-molybdenum cofactor (FeMoco), a cluster which acts as the active site of substrate binding and reduction in nitrogenase.[73]  Despite intense study, the identity of this atom could not be unambiguously established until the Fe VTC-XES study of Lancaster and co-workers.[74]  Likewise, the utility of this method has, on numerous occasions, been evidenced in recent catalysis research.  For example, Pushkar *et al.* demonstrated the feasibility of VTC-XES for detecting and probing the oxo bridges found in the $Mn_4Ca$ cluster of photosystem II, establishing a powerful tool for studying the O-O bond formation preceding $O_2$ evolution.[75]  Due to its increasing popularity, much research has been conducted to develop the theoretical underpinnings of VTC-XES and to identify spectral features that can serve as measures of various chemical parameters.  For example, a recent article by Pollock, *et al.*,[76] identifies a feature in the VTC-XES spectra of several $Fe-N_2$ complexes that can be attributed to a transition from the 2s2s σ* antibonding-orbital to the 1s core-hole.  The energy of this feature is then related to the N-N bond length and serves as a measure of the degree of activation of small molecules during catalytic reduction.[76] Finally, several review articles can be found that discuss VTC-XES in various levels of detail.[68-69, 77]

Here, a system of Zn compounds comprised of Zn metal, ZnO, and $ZnCl_2$ was chosen to reflect the feasibility of VTC-XES measurements with the present instrumentation along with its sensitivity to a variety of ligand environments. A similar study using an earlier, lower powered instrument investigated similar compounds and made a critical comparison across several theoretical treatments of VTC-XES.[37] As can be seen in Fig. 13, the present instrumental resolution clearly resolves the double-peak structure of the $K\beta_{2,5}$ lines. In addition, the $K\beta''$ transitions indicative of the ligand environment are clearly discernible for the oxide and chloride systems, with the former ~17 eV below the main peak, in rough agreement with values reported elsewhere for the relative $K\beta''$ position.[78] Finally, the non-resonant excitation process utilized with a broadband source again gives rise to multielectron features that can be observed toward high energies in the spectrum of metallic zinc. We note that similar XES features have been reported for other sample matrices elsewhere.[35, 79-80] However, it is interesting to note that while multielectron features were largely suppressed in the ligated Zn-compounds due to charge-transfer effects, this is expected not to be the case for early row transition metals whose properties are better described by the motif of Mott-insulators than charge-transfer semiconductors.[81]

Beyond investigations of the electronic details discussed so far, there exists a wealth of applications that would benefit from routine oxidation state analysis using laboratory-based XES. This has recently been demonstrated for hexavalent Cr identification using Cr Kα spectroscopy with the UW instrument,[40] and has also been used for identification of sulfur oxidation state in biochars[82] and phosphorus oxidation state in InP quantum dots[83-84] using a different very high-resolution lab-based XES system at UW.

Here, we consider whether benchtop XES can address the oxidation state of environmental arsenic. A pioneering work by Penrose found that of the two most common oxidation states, the trivalent species of arsenic is generally more toxic than the pentavalent.[85] XAFS techniques have emerged as essential alternatives for quantitative species fraction determinations of arsenic in solid matrices, such as in soils where the methodology can be paired with sequential extraction procedures[86] or HPLC-ICP-MS[8] to provide insights into the behavior of arsenic in ecological systems. Here, representative As(III) and As(V) compounds are presented as a demonstration study relevant for potential environmental speciation studies in a laboratory setting. For this study, samples were obtained in powder form and directly transferred to a polyimide pouch easily positioned in front of the source. As can be seen in Fig. 14a, the As Kα XES measurements reveal several noticeable spectral differences, including an energy shift that can be used as an indicator of oxidation state. Figure 14a also highlights the high intensity of these features, suggesting the potential of this technique in studies of dilute environmental samples. This approach to As speciation requires further investigation, as would other As fluorescence lines. We note that while the As Kα does have a clear energy shift, true environmental samples with As contamination also commonly have nontrivial Pb content, and that the Pb $L_{III}M_V$ emission line at 10551.6 eV can interfere with the As Kα XES.

Finally, we address XES of actinide materials. The L emission lines in U lie between 10 and 21 keV, with most toward the latter. Similar detector, source, and analyzer inefficiencies discussed above in the context of actinide XANES are problematic in U XES studies as well. In addition, the L-shell fluorescence lines more likely to be sensitive to chemical bonding are those involving shells closer to the valence and are consequently weaker transitions, again requiring instrumentation with minimal backgrounds. Nonetheless, U XES measurements of $(PPh_4)_2UCl_6$ and $(NMe_4)_2UO_2Cl_4$ are presented in Fig. 14b. The most prominent feature is the $L\beta_1$ ($L_{II}M_{IV}$) found around 17220 eV.[87] Little sensitivity to speciation was observed and while other L emission lines are observed in this energy region their inadequate separation from the tails of these features complicates their use as fingerprints for the relevant U species.

**V. Summary and Conclusions**

We present the instrumentation details and a wide variety of test study results for an improved laboratory spectrometer for XAFS and XES. This includes measurements that demonstrate important extremes for lab-based capability: EXAFS, VTC XES, and higher-energy performance. The assembled body of work using this new spectrometer, building on top of numerous studies by our research group[13, 35-41, 82-83, 88-89] and also ongoing research of several other research groups[23, 25-26, 34, 63, 90] strongly supports the position that laboratory XAFS and XES should not be judged in competition with synchrotron capability but should instead be appreciated for the new analytical capabilities that are enabled. These new capabilities hold high promise for routine materials analysis that can accelerate progress in electrical energy storage, coordination chemistry,[91] actinide chemistry,[63] and environmental and regulatory testing,[40] to name only a few prominent examples.

**Acknowledgements**

E. Jahrman and T. Fister were supported in part by the Joint Center for Energy Storage Research (JCESR), an Energy Innovation Hub funded by the U.S. Department of Energy, Office of Science, and Basic Energy Sciences, and by the U.S. Department of Energy through the Chemical Science and Engineering Division of Argonne National Laboratory. This material is based in part upon work supported by the State of Washington through the University of Washington Clean Energy Institute. The work of J. Rana and L. Piper was supported as part of the NorthEast Center for Chemical Energy Storage (NECCES), and Energy Frontier Research Center funded by the U.S. Department of Energy, Office of Science, Basic Energy Sciences under Award# DE-SC0012583. The work of N. Hyatt and M. Stennett was supported, in part by, the Nuclear Decommissioning Authority and EPSRC under grant numbers EP/N017617/1 and EP/R511754/1; and utilized the MIDAS facility at The University of Sheffield established with financial support from the Department for Business, Energy & Industrial Strategy. The authors would like to acknowledge the efforts of Carrie Siu and Dr. M. Stanley Whittingham in synthesizing the phase pure epsilon-VOPO4 and the efforts of Mateusz Zuba in collecting the synchrotron XAS data of that sample. R&D associated with the Los Alamos National Laboratory (LANL) spectrometer was funded under the Heavy Element Chemistry Program by the Division of Chemical Sciences, Geosciences, and Biosciences, Office of Basic Energy Sciences, U.S. Department of Energy and the U.S. Department of Energy. LANL is operated by Los Alamos National Security, LLC, for the National Nuclear Security Administration of U.S. Department of Energy (contract DE-AC52-06NA25396). Use of the Stanford Synchrotron Radiation Lightsource, SLAC National Accelerator Laboratory, was supported by the U.S. Department of Energy, Office of Science, Office of Basic Energy Sciences under Contract No. DE-AC02-76SF00515. This research used resources of the Advanced Photon Source, a U.S. Department of Energy (DOE) Office of Science User Facility operated for the DOE Office of Science by Argonne National Laboratory under Contract No. DE-AC02-06CH11357.

**Table 1**: Experimental details of XES and XAFS measurements performed in this work.  Acquisition times spanning multiple compounds refer to the time allotted to each sample.  Acquisition times reported in this table only includes the time required to span the energy range shown in the corresponding figure.  XAFS acquisition times are reported only for the transmission scans.

| Figure # | Anode | Power (W) | SBCA | Compound | Measurement | Acquisition Time (h) |
|---|---|---|---|---|---|---|
| 7 | W | 50 | Ge (422) | $\varepsilon$-VOPO$_4$ | XANES | 3.0 |
| 8a | Pd | 50 | Si (444) | NMC | XANES | 0.22 |
| 8b | W | 50 | Si (444) | NMC | XES K$\beta_{1,3}$ | 0.06 |
| 9a | W | 50 | Si (422) | CeO$_2$ | XANES | 1.0 |
| 9a | W | 50 | Si (422) | CePO$_4$ | XANES | 1.0 |
| 9b | Pd | 50 | Si (12,6,6) | (PPh$_4$)$_2$UCl$_6$ | XANES | 44 |
| 10 | W | 50 | Ge (422) | V | EXAFS | 3.4 |
| 11 | Pd | 100 | Si (551) | Ni | EXAFS | 6.9 |
| 12a | W | 50 | Ge (422) | $\varepsilon$-VOPO$_4$ | XES K$\beta$ | 12.4 |
| 12a | W | 50 | Ge (422) | V | XES K$\beta$ | 12.4 |
| 12a | W | 50 | Ge (422) | V$_2$O$_3$ | XES K$\beta$ | 12.4 |
| 12a | W | 50 | Ge (422) | VO$_2$ | XES K$\beta$ | 12.4 |
| 12a | W | 50 | Ge (422) | V$_2$O$_5$ | XES K$\beta$ | 12.4 |
| 12b | W | 50 | Ge (422) | V$_2$O$_3$ | XES K$\beta_{1,3}$ | 4.5 |
| 12b | W | 50 | Ge (422) | VO$_2$ | XES K$\beta_{1,3}$ | 4.5 |
| 12b | W | 50 | Ge (422) | V$_2$O$_5$ | XES K$\beta_{1,3}$ | 4.5 |
| 13 | Pd | 100 | Ge (555) | Zn | XES K$\beta_{2,5}$ | 9.6 |
| 13 | Pd | 100 | Ge (555) | ZnO | XES K$\beta_{2,5}$ | 8.0 |
| 13 | Pd | 100 | Ge (555) | ZnCl$_2$ | XES K$\beta_{2,5}$ | 11.5 |
| 14a | W | 50 | Si (555) | NaAsO$_2$ | XES K$\alpha$ | 1.4 |
| 14a | W | 50 | Si (555) | Na$_2$HAsO$_4\cdot$7H$_2$O | XES K$\alpha$ | 0.8 |
| 14b | Pd | 50 | Si (12,6,6) | (PPh$_4$)$_2$UCl$_6$ | XES L$\beta$ | 30 |
| 14b | Pd | 50 | Si (12,6,6) | TBA$_2$UO$_2$Cl$_4$ | XES L$\beta$ | 24 |

**Table 2:** Selected EXAFS fitting parameters for Ni foil measured at APS and at UW as compared to literature fits, x-ray diffraction (XRD), and neutron PDF analysis. Uncertainties correspond to one standard deviation.

|  | $S_o^2$ | R-factor | Shell1 | | Shell2 | | Shell3 | | Shell4 | |
|---|---|---|---|---|---|---|---|---|---|---|
|  |  |  | Ni-Ni (Å) | $\sigma^2$ ($10^{-4}$ Å$^2$) | Ni-Ni (Å) | $\sigma^2$ ($10^{-4}$ Å$^2$) | Ni-Ni (Å) | $\sigma^2$ ($10^{-4}$ Å$^2$) | Ni-Ni (Å) | $\sigma^2$ ($10^{-4}$ Å$^2$) |
| XRD[92] |  |  | 2.4863 |  | 3.5161 |  | 4.3063 |  | 4.9725 |  |
| Neutron PDF[93] |  |  | 2.487 (1) | 64 ± 1 |  |  |  |  |  |  |
| XAFS Lit.[94] |  |  | 2.493 (2) | 65 ± 2 |  |  |  |  |  |  |
| XAFS Lit.[93] | 0.84 (2) |  | 2.485 (2) | 64 ± 2 |  |  |  |  |  |  |
| APS 13-ID[47] | 0.90 (6) | 0.015 | 2.493 (4) | 67 ± 6 | 3.525 (5) | 96 ± 19 | 4.317 (6) | 91 ± 10 | 4.985 (7) | 79 ± 8 |
| UW | 0.81 (6) | 0.016 | 2.490 (4) | 61 ± 6 | 3.522 (5) | 76 ± 16 | 4.314 (7) | 92 ± 11 | 4.981 (8) | 79 ± 9 |

**Table 3:** Selected EXAFS fitting parameters for V foil measured at APS and at UW. Uncertainties correspond to one standard deviation. The V Foil synchrotron spectrum was taken from the XAFS spectra library maintained by the Center for Advanced Radiation Sources at APS.

|  | $S_o^2$ | R-factor | Shell1 | | Shell2 | | Shell3 | | Shell4 | | Shell5 | |
|---|---|---|---|---|---|---|---|---|---|---|---|---|
|  |  |  | Ni-Ni (Å) | $\sigma^2$ ($10^{-4}$ Å$^2$) | Ni-Ni (Å) | $\sigma^2$ ($10^{-4}$ Å$^2$) | Ni-Ni (Å) | $\sigma^2$ ($10^{-4}$ Å$^2$) | Ni-Ni (Å) | $\sigma^2$ ($10^{-4}$ Å$^2$) | Ni-Ni (Å) | $\sigma^2$ ($10^{-4}$ Å$^2$) |
| APS 13-ID | 0.70 (7) | 0.017 | 2.608 (7) | 81 ± 13 | 3.012 (8) | 72 ± 15 | 4.259 (11) | 120 ± 30 | 4.994 (13) | 150 ± 50 | 5.217 (13) | 80 ± 30 |
| UW | 0.76 (7) | 0.013 | 2.620 (6) | 81 ± 12 | 3.025 (7) | 86 ± 14 | 4.279 (10) | 140 ± 30 | 5.017 (12) | 120 ± 30 | 5.240 (12) | 90 ± 30 |

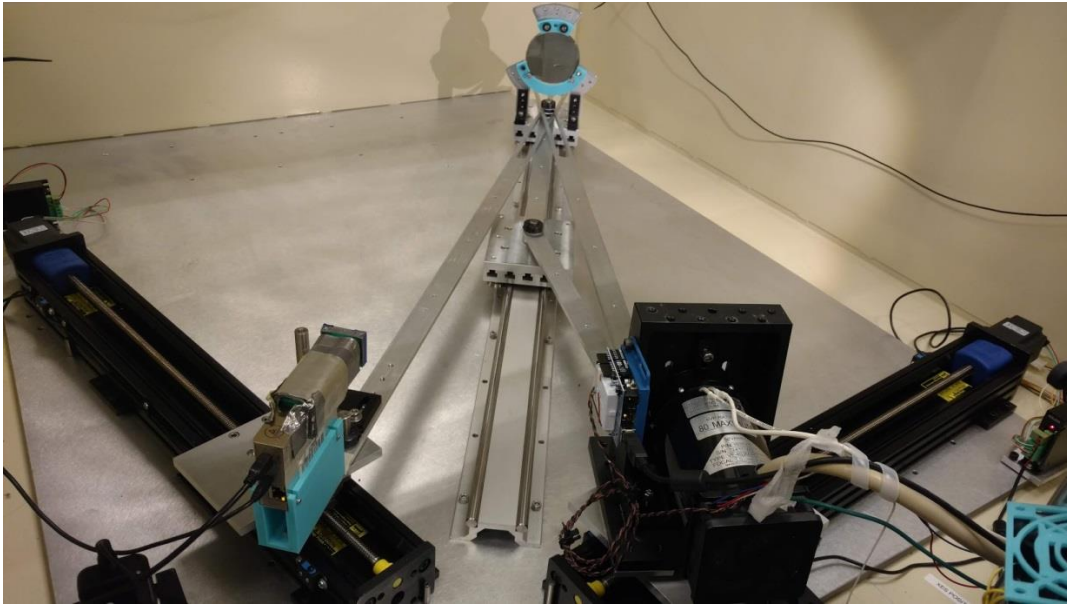

**Figure 1**: Corner perspective of spectrometer in XANES configuration. The SBCA and source are mechanically coupled to the center carriage. The two-axis tilt is no longer utilized. Source and detector are at $\alpha = 40°$ (see Fig. 2).

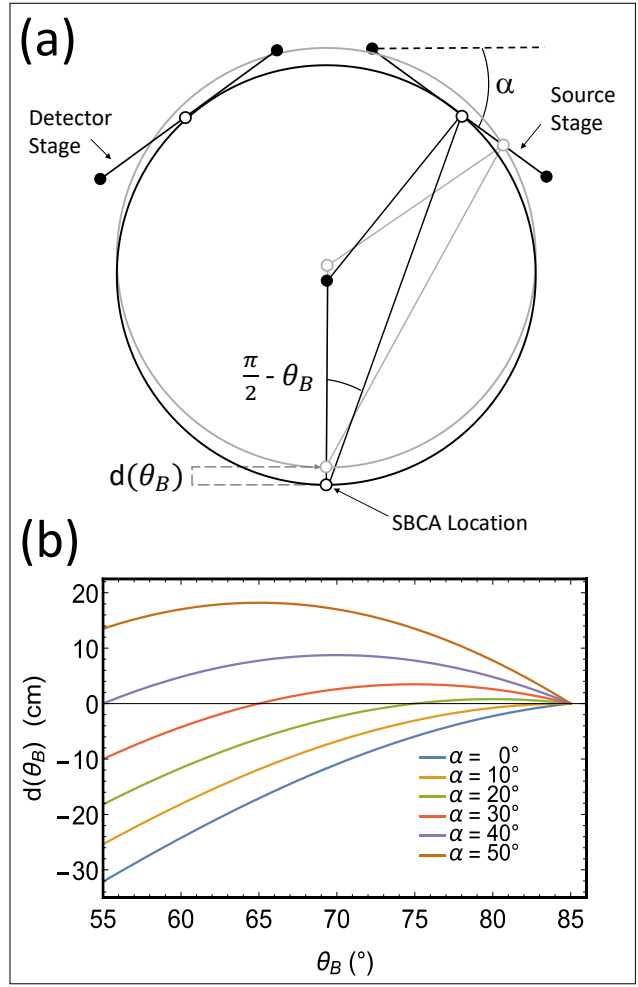

**Figure 2**: (a) Illustration depicting the parameter α and $\theta_B$. The SBCA resides at the bottom of the Rowland circle while the carriage coupling the SBCA location and the source as represented by the hollow dot is at the center of the Rowland circle. The diagonal line represents the travel range of the source with dots at its end points. (b) The magnitude of the SBCA's displacement from its location, d($\theta_B$), at $\theta_B = 85°$ is plotted as a function of $\theta_B$ for various values of α.

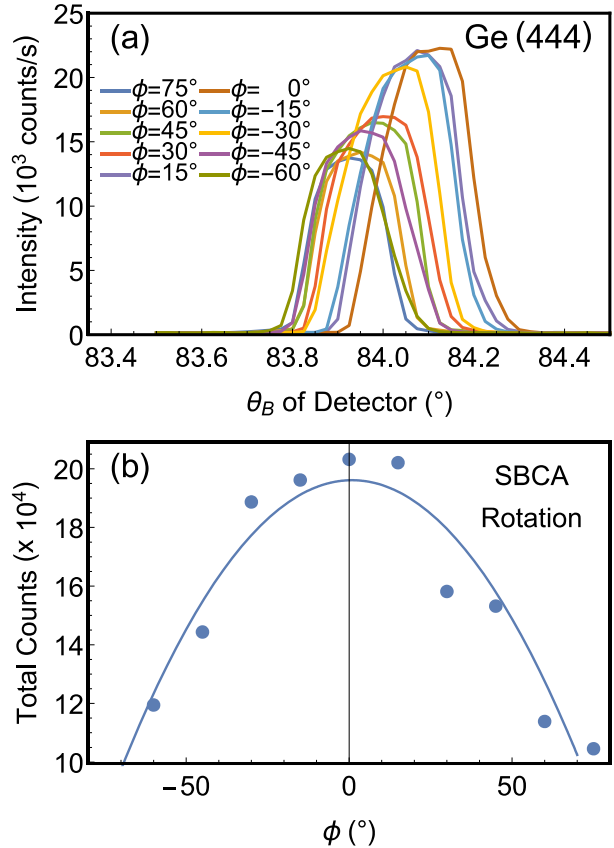

**Figure 3**: (a) While holding the source fixed at $\theta_B = 84°$, the detector was scanned from 83.5 to 84.5°. Scans were taken at various rotations of the SBCA about its center with the optimum position designated as 0°. Data was taken off the 444 harmonic of a Ge SBCA using a x-ray tube source with a Pd anode operated at 52.5 W tube power. The duration of each scan was approximately 45 seconds. (b) The total number of counts, as integrated over the range from 83.5 to 84.5° for each scan, is shown as a function of the analyzer's angular rotation. The solid line is a quadratic fit.

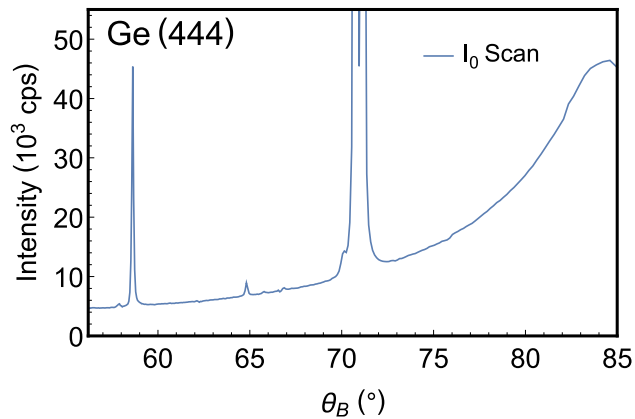

**Figure 4**: An $I_0$ scan spanning the entire range of the instrument. Data was collected using the 444 harmonic of a Ge SBCA. An x-ray tube source with a Pd anode was operated at 100 W power. Fluorescence lines can be seen from Cu Kα and Kβ lines as well as a small W line around 8400 eV. This last line is likely due to some small number of W atoms from the filament being deposited onto the surface of the target anode as has been discussed elsewhere.[31]

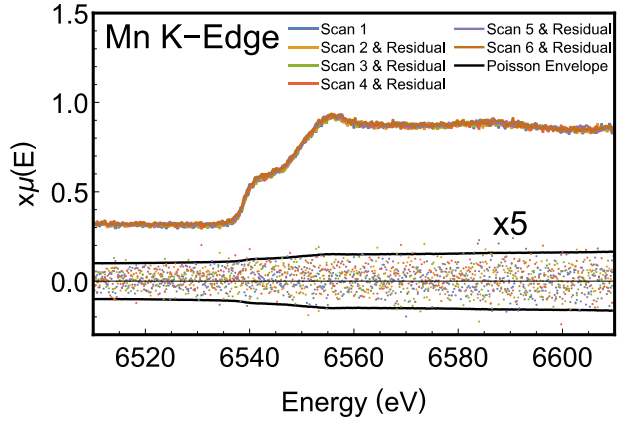

**Figure 5**: Six consecutive scans are shown of a transmission mode measurement across the K-edge of the Mn foil. Measurements were collected using a Si (440) SBCA. An x-ray tube source with a W anode was operated at 25 W with a 10 µm thick Zn foil acting as an absorber to suppress the W fluorescence line observed on the Si (660) harmonic in accordance with methods previously reported, although done here without a slit system.[95] The residuals between subsequent scans are shown at the bottom of the figure (magnified five times) with a Poisson envelope enclosing two standard deviations.

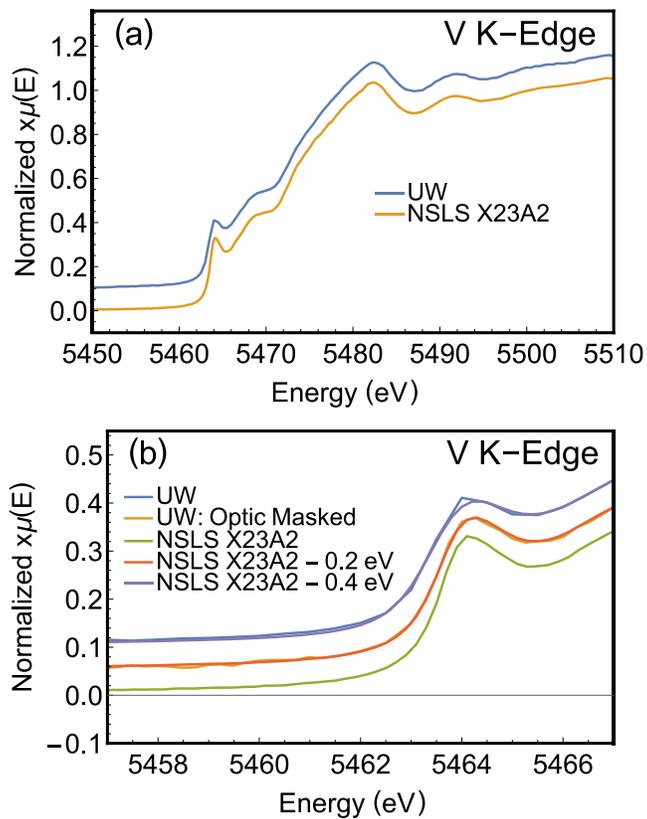

**Figure 6**: (a) XANES spectra of the V foil collected using an x-ray tube source with a W anode and operated at 50 W power. Comparison was made to synchrotron results and offset for clarity, see the text for discussion. (b) An enlarged view of the pre-edge feature at ~5464 eV including comparison with synchrotron results with the indicated Gaussian broadening. Laboratory-based measurements used either a masked or unmasked Ge (422) SBCA. Spectra are offset for clarity.

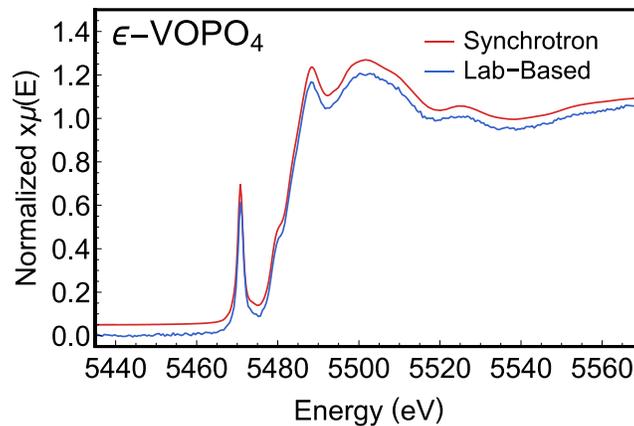

**Figure 7**: The V K-edge XANES spectra of a vanadyl phosphate-based battery laminate. Spectra were acquired with the present instrumentation (Lab-Based) and at APS 9-BM (Synchrotron). The spectra are offset for clarity of presentation. The full range of scans was chosen to extend from 5390 eV out to 10 Å$^{-1}$ to ensure proper normalization and background removal for comparison to the synchrotron.

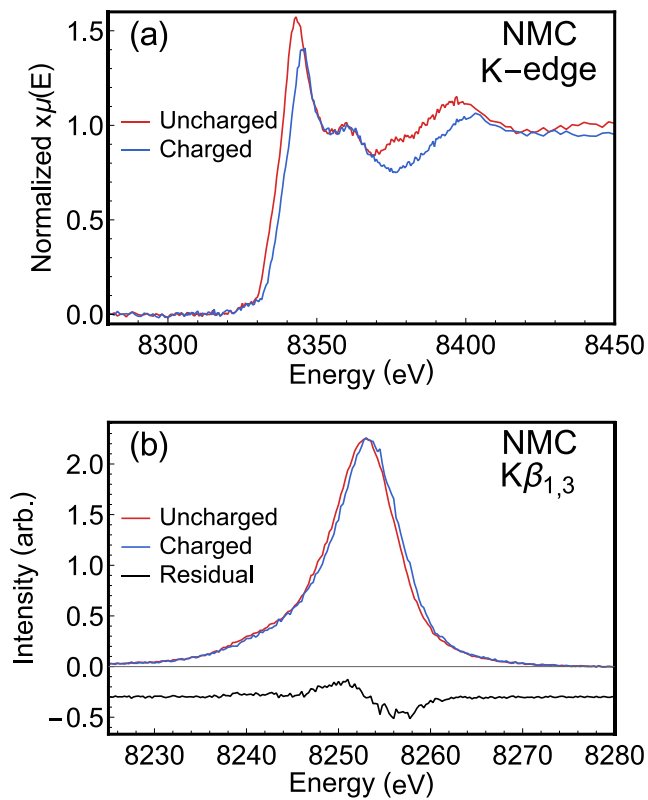

**Figure 8**: (a) XANES spectra of uncharged and charged battery laminates of NMC composition. Data was acquired out to 10 Å$^{-1}$ to ensure proper normalization and data was collected at lower energies to aid background removal. (b) XES spectra of a charged and uncharged NMC laminate. The residual of the two spectra is displaced below the main results. Peak count rates were around 12,000 counts per second for the uncharged laminate and 6,000 for the charged laminate.

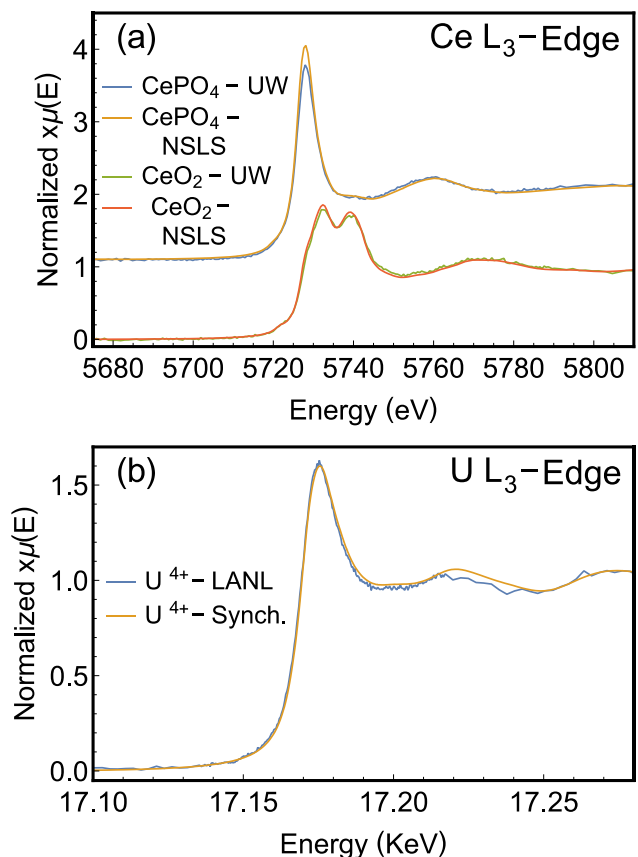

**Figure 9**: (a) XANES spectra of $CePO_4$ and $CeO_2$, representative $Ce^{3+}$ and $Ce^{4+}$ compounds, respectively. Reference spectra were acquired on beamline X23-A2 of the National Synchrotron Light Source (NSLS). (b) Comparison of synchrotron (endstation 11-2 at SSRL) and lab-based U $L_3$-edge XANES for $(PPh_4)_2UCl_6$, a $U^{4+}$ reference compound. Data was calibrated to the maximum of the first derivative of the K-edge spectrum of a yttrium foil at 17038.4 eV.

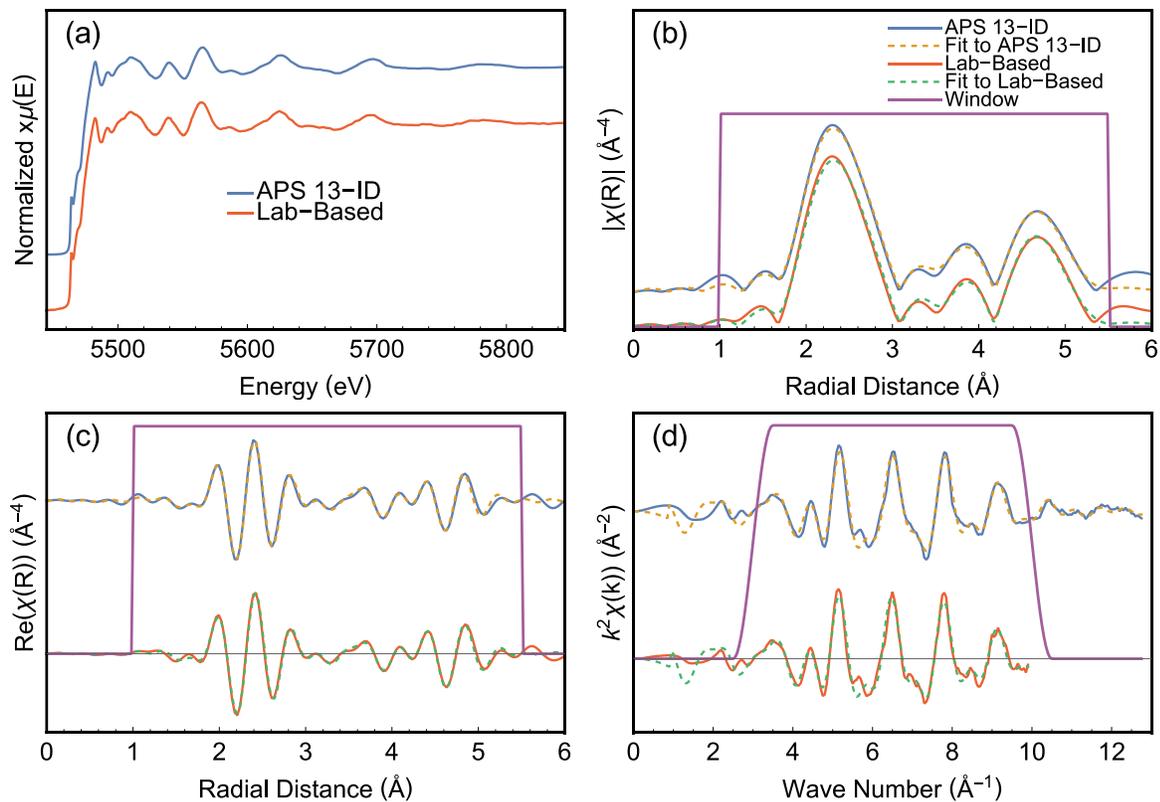

**Figure 10:** EXAFS of V Foil collected at UW (Lab-based) compared to synchrotron results (APS 13-ID). Results are shown in (a) energy space, (b) along with the magnitude of the EXAFS Fourier transform, (c) the real part of the EXAFS in radial space, and (d) the EXAFS with quadratic weighting in k-space, respectively. Also shown are the fitted models, see the text for discussion.

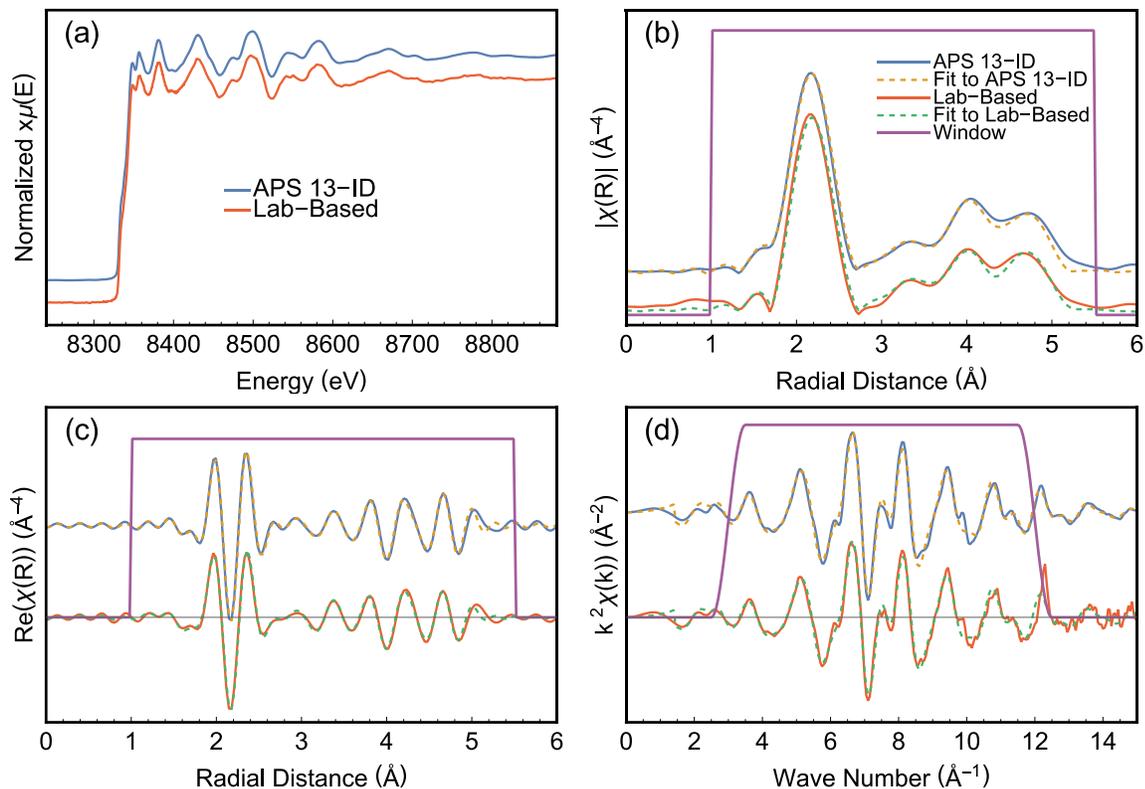

**Figure 11:** EXAFS of Ni Foil collected at UW (Lab-based) compared to synchrotron results (APS 13-ID). Results are shown in energy space (a), along with the magnitude of the EXAFS in radial space (b), the real part of the EXAFS in radial space (c), and the EXAFS with quadratic weighting in k-space (d), respectively. Also shown are the fitted models acquired from Artemis.[47] Data was collected using 100 W power for a Pd anode x-ray tube and using a Si (551) SBCA. Measurement times were 1.7 and 6.9 h for $I_0$ and $I_T$, respectively.

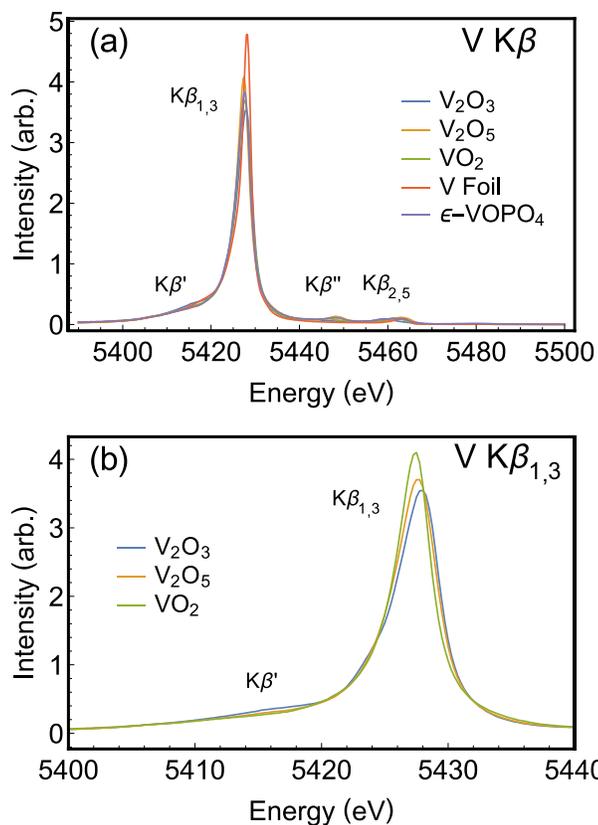

**Figure 12**: (a) The full range of V Kβ XES from a collection of V compounds measured in the lab spectrometer. Measurement times were 12.4 h for all samples. Note that the vanadyl phosphate data represents three scan ranges, with the main scans spanning 5395 eV to 5485 eV, this range was supplemented joined with supplemental data sets to span the entire range shown and permit equivalent background subtractions for all systems. (b) V Kβ$_{1,3}$ XES from a suite of oxides measured in the lab spectrometer.

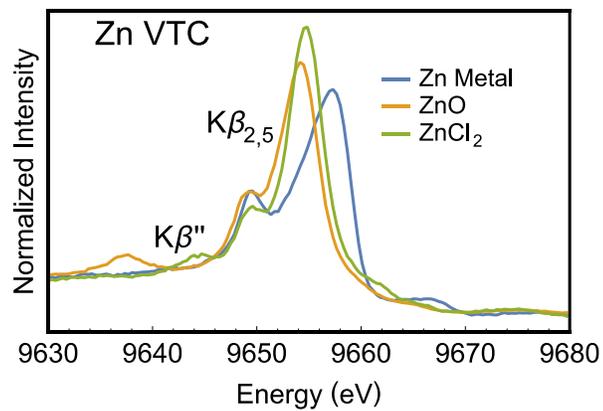

**Figure 13**: VTC-XES spectra of Zn metal, ZnO, and ZnCl$_2$ after background subtraction and integral normalization across the full VTC energy range.

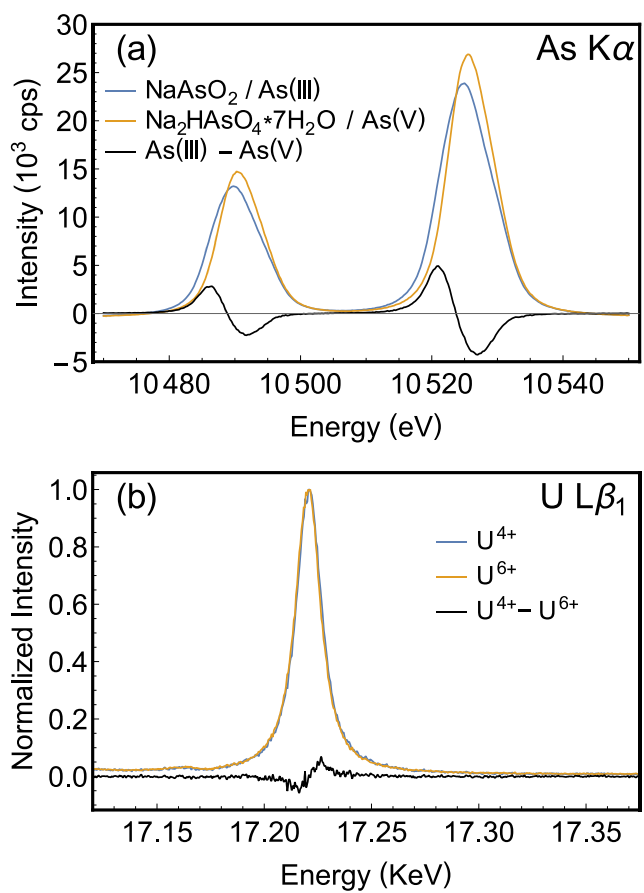

**Figure 14**: (a) The As Kα XES spectra of trivalent and pentavalent arsenic oxide species (a). The intensity scale is for the NaAsO$_2$ sample; the intensity of the Na$_2$HAsO$_4$*7H$_2$0 has been scaled upward by ~30% to give it the same integral intensity for ease of comparison for the energy shift as a function of As oxidation state. The study used a Si (555) toroidally bent crystal analyzer following an in-house design.[41] (b) Collected Lβ$_1$ XES spectra of (PPh$_4$)$_2$UCl6 and (NMe$_4$)$_2$UO$_2$Cl$_4$, which are in the U$^{4+}$ and U$^{6+}$ state, respectively. The most intense spectral feature is the Lβ$_1$, though less intense features can be found toward lower energies. The spectra are peak normalized here for comparison. No change in spectrum was observed across any of the scans, indicating no radiation damage. The data was calibrated to the maximum of the Kα of a Mo foil at 17480 eV.